\documentclass[runningheads]{llncs}
\usepackage{graphicx}
\usepackage{epsfig}
\usepackage{dsfont}
\usepackage{amsmath}
\usepackage{amssymb}
\usepackage{color}
\usepackage{boldline,multirow}
\usepackage{siunitx}
\usepackage{booktabs}
\usepackage{soul}
\usepackage{cite}
\usepackage{cases}
\usepackage{hyperref}
\usepackage{bbding}
\newcommand{\ucite}[1]{\cite{#1}} 
\begin{document}
\title{SATr: Slice Attention with Transformer for Universal Lesion Detection}
\titlerunning{SATr: Slice Attention with Transformer for Universal Lesion Detection}

%

\author{Han Li\inst{1,2} \and Long Chen\inst{2,3} \and Hu Han\inst{2}\Envelope \and S. Kevin Zhou\inst{1,2}\Envelope}
\institute{1. School of Biomedical Engineering \& Suzhou Institute for Advanced Research
Center for Medical Imaging, Robotics, and Analytic Computing \& LEarning (MIRACLE)
University of Science and Technology of China, Suzhou 215123, China\\ 2. Key Lab of Intelligent Information Processing of Chinese Academy of Sciences (CAS),
Institute of Computing Technology, CAS, Beijing, 100190, China\\ 3. School of Computer Science and Technology, University  of  the  Chinese  Academy of  Science\\
\email{\{han.li,long.chen\}@miracle.ict.ac.cn, hanhu@ict.ac.cn}}

\authorrunning{H.~Li, et.al}



\maketitle              
\begin{abstract}
  Universal Lesion Detection (ULD) in computed tomography plays an essential role in computer-aided diagnosis.  Promising ULD results have been reported by multi-slice-input detection approaches which model 3D context from multiple adjacent CT slices, but such methods still experience difficulty in obtaining a global representation among different slices and within each individual slice since they only use convolution-based fusion operations.
  In this paper, we propose a novel Slice Attention Transformer (SATr) block which can be easily plugged into convolution-based ULD backbones to form hybrid network structures. Such newly formed hybrid backbones can better model long-distance feature dependency via the cascaded self-attention modules in the Transformer block while still holding a strong power of modeling local features with the convolutional operations in the original backbone.  Experiments with five state-of-the-art methods show that the proposed SATr block can provide an almost free boost to lesion detection accuracy without extra hyperparameters or special network designs. 

\keywords{Universal lesion detection \and Slice Attention \and  Transformer.}
\end{abstract}

\section{Introduction}\label{sec:introduction}
Universal Lesion Detection (ULD)  in computed tomography (CT) \ucite{zlocha2019one-stage,tao2019improving,zhang2019anchor_free,zhang2020Agg_Fas,tang2019uldor,yan20183DCE,li2019mvp,yan2019mulan,yang2020alignshift,cai2020deep,li2020bounding,zhang2020revisiting,yan2020learning, cai2021deep,tang2021weakly,yang2021asymmetric,li2021conditional,lyu2021segmentation}, aiming to localize different types of lesions instead of identifying lesion types \ucite{boot2020diagnostic,yu2020deep,ren2021retina, baumgartner2021nndetection,shahroudnejad2021tun,luo2021oxnet,chen2021ellipsenet,yang2021leveraging,lin2021automated,zhao2021positive}, plays an essential role in computer-aided  diagnosis (CAD)\cite{zhou2021review,zhou2019handbook}. ULD is a challenging task because different lesions have diverse shapes and sizes, easily leading to false positive and false negative detections. Mainly inspired by the clinical fact that radiologists need several adjacent slices for locating and diagnosing lesions on one CT slice, most existing ULD methods take several adjacent 2D CT slices 
as the inputs to a 2D network architecture  \ucite{zhang2019anchor_free,zhang2020Agg_Fas,yan20183DCE,li2019mvp,yan2019mulan,yang2020alignshift,cai2020deep,zhang2020revisiting,tang2021weakly,yang2021asymmetric,li2021conditional,lyu2021segmentation} or directly adopt 3D  network designs \cite{cai2020deep}  that take 3D volume as input to extract more 3D context information. 
While both 2D and 3D methods have yielded great ULD performances, the multi-slice-input based 2D detection methods are much more popular than pure 3D fashion because 2D networks benefit from robust 2D models pretrained from large-scale data whereas publicly available 3D medical datasets are not large enough for robust 3D pretraining.

\begin{figure}[t]
\centering
\setlength{\abovecaptionskip}{0.cm}

\includegraphics[scale=0.4]{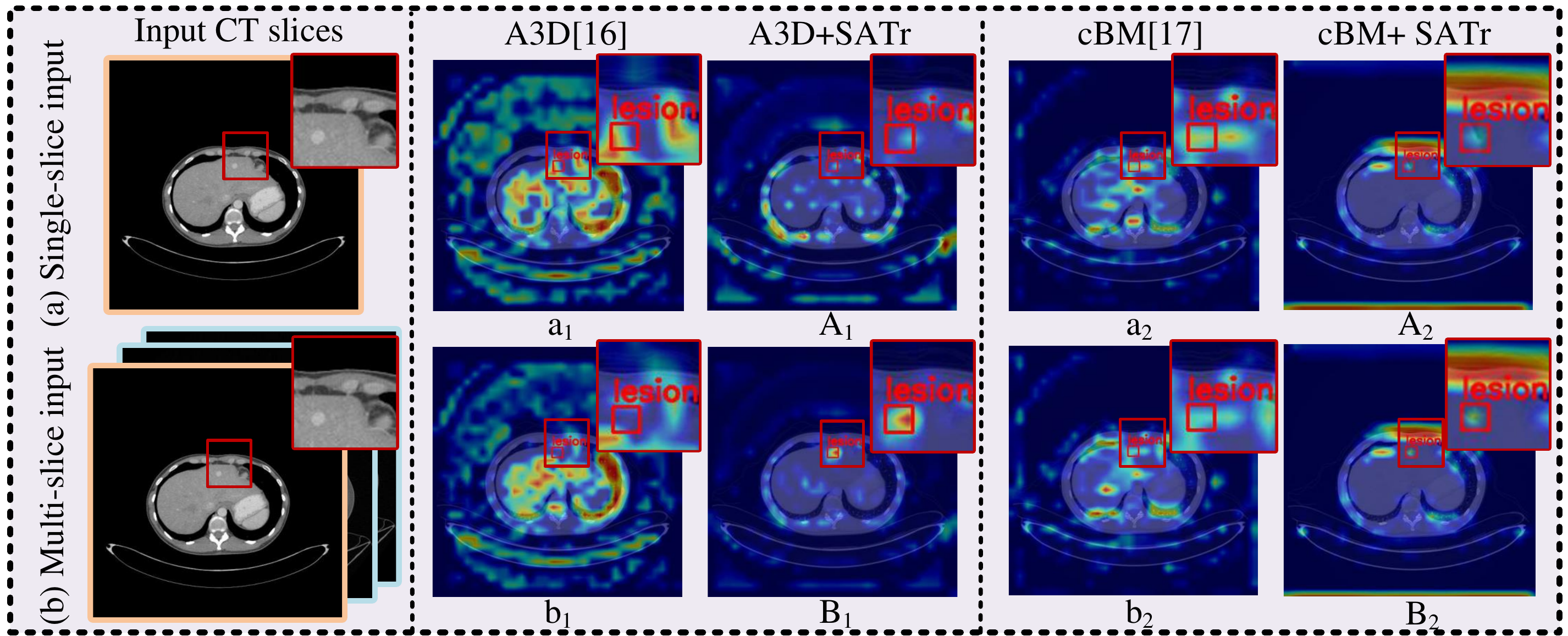}
\caption{Visualization of the CAMs of two state-of-the-art ULD methods with and without using our SATr block under (a) single-slice-input and (b) multi-slice-input scenarios. }
\label{fig:fig1_CAM}
\end{figure}

While achieving success in ULD, the multi-slice-input based 2D approaches have inherent limitations:
(i) \emph{Weak global context modeling within each slice.}
Most existing ULD methods are fueled by the successful Convolutional Neural Networks (CNN) pretrained backbones (e.g., DenseNet\cite{huang2017densely}) which learn local features via convolution operations in a hierarchical fashion (e.g., FPN\cite{lin2017fpn}) as powerful image representations. Despite the strong power of local feature extraction, CNNs show their weakness in dealing with global (long-distance spatial) contexts \cite{peng2021conformer}. As shown in Fig.\ref{fig:fig1_CAM} (a), the single-slice CAMs of two CNN-based ULD backbones contain flocks of redundant and inaccurate activation regions which is due to the limitation of CNN backbones in handling global contexts within each slice for ULD tasks.   
(ii) \emph{Weak inter-slice context modeling.}
When dealing with the multi-slice input, most ULD methods \ucite{zhang2020Agg_Fas,yan20183DCE,li2019mvp,yan2019mulan,cai2020deep,zhang2020revisiting,tang2021weakly,lyu2021segmentation} adopt independent 2D convolutional operations for each slice to extract independent features, and the independent features further go through a 3D convolution for feature fusion. Such convolution-based fusion methods are good at dealing with local features, but they unfortunately deteriorate in handling global features among different slices. To tackle this, some ULD approaches \ucite{yang2020alignshift,yang2021asymmetric,li2021conditional} propose to reshuffle feature channels among different slices which relieves this issue to some degree. But their ability in capturing rich global representations are still weak due to the use of pure convolutional operations. As shown in  the CAMs in Fig.\ref{fig:fig1_CAM}, with the help of convolution based multi-slice fusion, the number of redundant and inaccurate activation regions can be decreased than single-slice-input based methods; however, these methods still cannot capture rich global contextual information due to the limit of CNN receptive field. 

Recently, CNN-transformer hybrid architectures  have been introduced to several visual tasks\ucite{peng2021conformer,xu2021vitae,mao2021dual}. Thanks to the Multihead Self-Attention (MSA) mechanism, Multilayer Perceptron (MLP) structure and CNN-transformer hybrid design can reflect the complex spatial transforms for both global and local feature dependency. Unfortunately, such hybrid architectures are designed for single-image-input (or single-slice-input) cases, and thus may not work well in the multi-slice-input scenarios. The main reason is that they may unavoidably introduce a lot of redundant information from adjacent slices (e.g., non-lesion slices or too-far slices), which may be harmful for ULD. Equally using features from every slice and feeding them to the naive transformer block will largely diminish the representation of the key CT slice.

To address the above issues, we hereby propose a novel Slice Attention Transformer (SATr) block, which works properly with multi-slice-input and assists can effectively extract the global representations from both each individual slice and multiple adjacent slices. The SATr block can be easily plugged into the popular convolution-based ULD detection backbones to form hybrid network structures, thus reaping the advantages from both Transformer and CNN worlds.

To validate the effectiveness of our method, we conduct extensive experiments on the DeepLesion dataset \cite{yan18deeplesion} based on five SOTA ULD methods under all training data settings, and also test the SATr with less training data (25\% and 50\%)  based on two SOTA ULD methods.

\begin{figure}[t]
\centering
\setlength{\abovecaptionskip}{0.cm}

\includegraphics[scale=0.21]{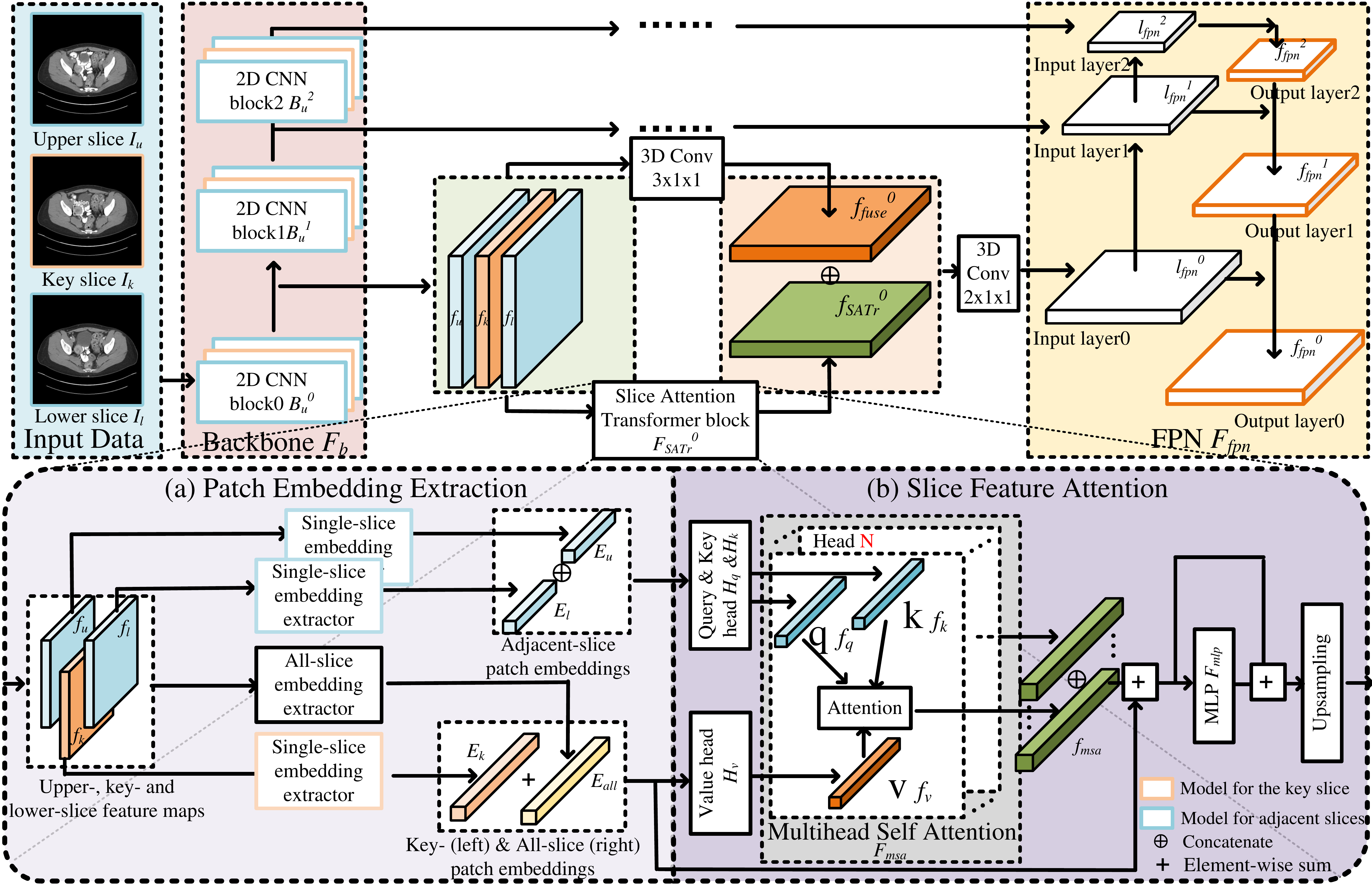}
\caption{The network architecture of ULD with the proposed SATr block.}

\label{fig:fig2_network_architecture}
\end{figure}

\section{Method}

As shown in Fig. \ref{fig:fig2_network_architecture},
The SATr block features two novel modifications of a naive transformer block: \emph{i) Enhancement of value vector with key-slice feature.} As described above, arbitrarily using the all-slice feature causes an opposite effect. Hence we separate the key-slice feature and add it up with the all-slice feature to serve as the value vector. In this way, the key-slice feature is largely preserved and SATr applies more strength to catching global cues between the key-slice's feature and other slices' feature. It should be noted that the value vector also entirely contains all-slice feature, therefore SATr is still highly effective in modeling global contexts both within each slice and among different slices.
\emph{ii) Removal of key-slice feature from the query and key vectors.} Experiments show that when using the all-slice feature to generate query and key vectors, SATr is prone to catching feature dependency within the key-slice feature itself (network lazy or overfittin), this is reasonable because the key-slice feature itself should be the main contributor for key-slice lesion detection task, but it completely conflicts with the motivation of introducing multiple adjacent slices to assist ULD. Admittedly, this removal sacrifices the feature dependency learned within the key slice, but it forces the SATr to learn more dependency to better make up the weakness of CNN backbone.

We hereby take the three-slice-input ULD method as a working example to illustrate our method while most SOTA results are reached under 7-slice or 9-slice settings. The SATr blocks are inserted between the feature extractor block (i.e., blocks in backbones) and each feature collector (e.g., FPN) without extra modifications on the original network. Section \ref{sec:backbone} details the common backbone of  multi-slice-input ULD methods; Section \ref{sec:SATr} explains the newly introduced SATr block, and Section \ref{sec:hybrid} introduces the hybrid network with SATr blocks.

\subsection{Multi-slice-input backbone}\label{sec:backbone}
In the multi-slice-input fashion, the ULD method is trained to localize the lesion in the key slice $I_k$, while the adjacent slices, including the same number of upper slices, $I_u=[I_u^1, ..., I_u^N]$, and lower slices, $I_u=[I_l^1, ..., I_l^N]$, are used to assist the lesion detection for the key slice. Without loss of generality, we set $N=1$. 
Hence, the input data can be formulated as $I=[I_u,I_k,I_l] \in \mathcal{R}^ {3 \times 1 \times W \times H}$, where $W$ and $H$ are the width and height of the input CT slices, respectively.

Similar to most backbones, the multi-slice-input backbone $F_b$ consists of several continuous CNN blocks $B=[B^{0},..., B^{M}]$ (e.g, Dense blocks in DenseNet), and further each block utilizes several separate sub CNN blocks $B^{m}= [B_u^{m}, B_k^{m}, B_l^{m}]$ to deal with the multi-slice input; thus the features from different slices $f=[f_u,f_k,f_l] $ are independent after these blocks. Specifically,
\begin{equation}
  f^{m}=[f_u^{m}, f_k^{m},f_l^{m}],~f^{m+1}=[f_u^{m+1}, f_k^{m+1},f_l^{m+1}]=[B_u^{m}(f_u^m),B_k^{m}(f_k^m),B_l^{m}(f_l^m)],
\end{equation}
where $B_k^m$ is the sub-CNN block for the key-slice feature in the $m$-th CNN block $B^m$, and $f^{m} \in \mathcal{R}^ {3\times C^m \times \frac{W}{R^m}\times \frac{H}{R^m}} $(or $f^{m+1}$) is the input (or the output) of $B^m$. $R^m$ and $C^m$ are the downsampling ratio and channel number of CNN block $B^m$.

Afterward, the output feature $f^{m} \in \mathcal{R}^ {3\times C^m \times \frac{W}{R^m}\times \frac{H}{R^m}}$ is fed into a 3D convolution $Conv_{3d}$ with $3\times1\times1$ Kernal to fuse the feature among different slices:
\begin{equation}
 f_{fuse}^{m}=Unsqueeze(Conv_{3d}(f^{m}),0),~~
 f_{fuse}=[f_{fuse}^{0},..., f_{fuse}^{M}],
\end{equation}
where $f_{fuse}^m \in \mathcal{R}^ {C^m \times \frac{W}{R^m}\times \frac{H}{R^m}}$ acts as the  $m$-th input layer $l_{fpn}^m$ in FPN $F_{fpn}$:

\begin{equation}
 l_{fpn}^m=f_{fuse}^m,~~
 f_{fpn}= F_{fpn}(l_{fpn}^0, ...,l_{fpn}^M).
\end{equation}
The $f_{fpn}$ is the final output of the FPN $F_{fpn}$.

\subsection{Slice attention transformer}\label{sec:SATr}
As shown in Fig. \ref{fig:fig2_network_architecture}, the SATr block contains two stages, i.e, patch embedding extraction and slice feature attention.

\begin{figure}[t]
\centering
\setlength{\abovecaptionskip}{0.cm}

\includegraphics[scale=0.31]{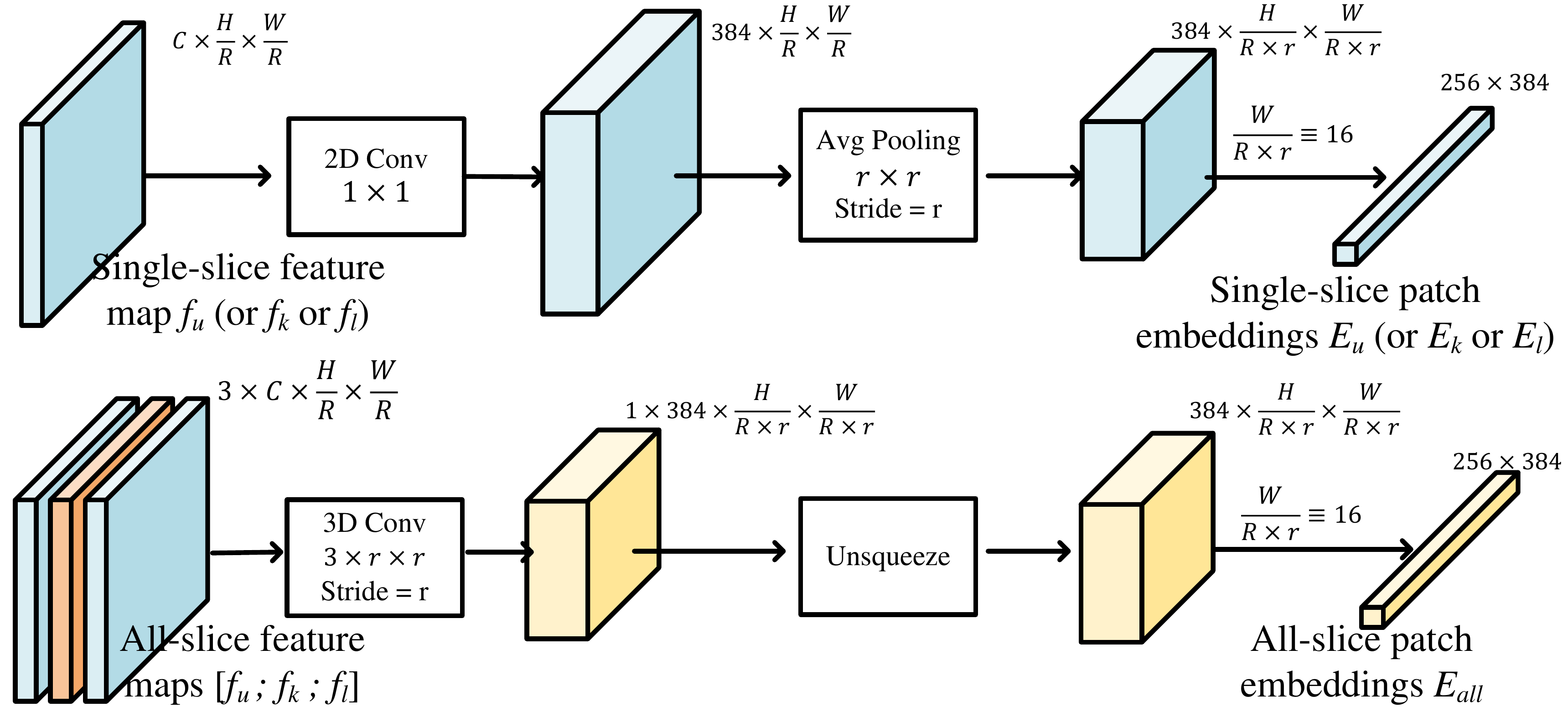}
\caption{The architecture of the proposed single-slice patch embedding extractor (upper) and all-slice patch embedding extractor (lower).}

\label{fig:fig3_extractor}
\end{figure}

\textbf{Patch embedding extraction.} Motivated by\ucite{peng2021conformer,zhu2020deformable,carion2020end}, as shown in Fig.\ref{fig:fig3_extractor}, we use the `Convolutional $+$ pooling' manner for single-slice patch embedding extraction. Considering that the extraction is the same among different slices and different CNN blocks $B^m$, we only formulate the extraction processes of key-slice $f_k \in \mathcal{R}^ {C \times \frac{W}{R}\times \frac{H}{R}}$ and remove the CNN block index $m$ in this section:

\begin{equation}
 f_k^E=AP(Conv_{2d}(f_k),r),~~ r= \frac{W}{16R}= \frac{H}{16R},
\end{equation}
where $Conv_{2d}$ denotes a 2D convolution with $Kernel=(1 \times 1)$ and $384$ output channels, the $AP$ is the average pooling operation with pooling size $r$ and stride $r$. We change the pooling size accordingly to get the fixed size output feature $f^e=[f_u^E,f_k^E,f_l^E]\in \mathcal{R}^ {3\times 384 \times 16 \times 16}$. Now the size of each slice's patch embeddings $[E_u, E_k, E_l] \in \mathcal{R}^ {3\times 256 \times 384}$ is also fixed after feature reshaping.

As for the all-slice patch embedding extraction, we utilize a 3D convolution $Conv_{3d}$ to fuse features among different slices which is the same with the fusion manner in the backbone.
\begin{equation}
\begin{aligned}
 f_{all}^E&=Conv_{3d}(f).
\end{aligned}
\end{equation}
Differently, the  $Conv_{3d}$ is with  $Kernal=(3\times r \times r)$  and $stride=r$, so the $f_{all}^E$ shares the same shape with single-slice patch embeddings, and the all-slice patch embeddings $E_{all} \in \mathcal{R}^ {256 \times 384}$ can be also generated after feature reshaping.

It is worth noting that all patch embeddings, including single- and all-slice embeddings, are with the same spacial size to fit the followed transformer block.

\textbf{Slice feature attention.}
Within the slice feature attention, we still follow the `$<q,k,v>$ head-MSA-MLP' design as in the naive transformer block, but we modify the input of $q\&k\&v$ according to the ULD multi-slice-input scene.

As for the query head $H_q$ and key head $H_k$, we concatenate the patch embedding of all adjacent slices ($E_u \& E_l$) to act as the input:

\begin{equation}
\begin{aligned}
 f_q=H_q([E_u;E_l]),
 f_k=H_k([E_u;E_l]),
\end{aligned}
\end{equation}
where $f_q$ and $f_k$ are the q value and k value, respectively. As for the value head $H_v$, the sum of key-slice and all-slice patch embedding is the input:

\begin{equation}
\begin{aligned}
 f_v=H_v(E_{all} + E_{k}),
\end{aligned}
\end{equation}
where $f_v$ is the v value and `$+$' denotes the element-wise sum.

The afterward design follows the naive Transformer block including MSA $F_{msa}$ and MLP $F_{mlp}$:
\begin{equation}
f_{msa}=F_{msa}(f_q,f_k,f_v),~~
f_{mlp}=F_{mlp}(f_{msa}+ f_v),~~
f_{tr}=f_{mlp}+f_{msa}.
\end{equation}
The generated output feature $f_{tr}\in \mathcal{R}^ {256 \times 384}$ will be interpolated and reshaped to the original size to get the final output of our proposed SATr,

\begin{equation}
\begin{aligned}
f_{SATr}&=ReShape(Interpolate(f_{tr})),
\end{aligned}
\end{equation}
where $f_{SATr} \in \mathcal{R}^ { C^m \times \frac{W}{R^m}\times \frac{H}{R^m}}$ is the output of the SATr. 

\subsection{Hybrid network}\label{sec:hybrid}
The proposed SATr $F_{SATr}$ can be easily implemented into convolution-based detection backbones to form hybrid network structures:
\begin{equation}
 {f}_{SATr}^{m}=F_{SATr}^m(f^{m}),~~
 \hat{f}_{fuse}=Conv_{3d}([f_{SATr}^{m};f_{fuse}^{m}]),
\end{equation}
where  $Conv_{3d}$ denotes a 3D convolution with $Kernal=(2 \times 1 \times 1)$, the outputs of `SATr' $f_{SATr}^{m}$ and the original backbone $f_{fuse}^{m} $will be concatenated and further fused by $Conv_{3d}$ to serve as the $m$-th input layer $l_{fpn}^m$ of FPN:
\begin{equation}
l_{fpn}^m=\hat{f}_{fuse}^m,~~
f_{fpn}= F_{fpn}(l_{fpn}^0, ...,l_{fpn}^M).
\end{equation}

\section{Experiments}
\subsection{Dataset and setting}
Our experiments are conducted on the ULD dataset DeepLesion\cite{yan18deeplesion}.  The dataset contains 32,735 lesions on 32,120 axial slices from 10,594 CT studies of 4,427 unique patients. Most existing datasets typically focus on one type of lesion, while DeepLesion contains a variety of lesions with large diameter range (from 0.21 to 342.5mm). The 12-bit intensity CT is rescaled to [0,255] with different window range settings used in different frameworks. Also, every CT slice is resized and interpolated according to the detection frameworks' setting. We follow the official split, i.e., $70\%$ for training, $15\%$ for validation and $15\%$ for testing. To further test our method's performance on a small dataset, we also conduct experiments based on  25\% and 50\% training data.  The number of false positives per image (FPPI) is used as the evaluation metric. For training, we use the original network architecture and settings.
\newsavebox{\tablebox}
\begin{table}[t]
\centering
\caption{Sensitivity (\%) at various FPPI under full training dataset settings on the testing dataset of DeepLesion \cite{yan18deeplesion}.} \label{results}
\begin{lrbox}{\tablebox}
\begin{tabular}{p{40mm}p{13mm}<{\centering}p{12mm}<{\centering}p{23mm}p{23mm}p{23mm}p{23mm}p{23mm}}
\toprule[1.5pt]
\textbf{Methods}&\textbf{data}&\textbf{slices}&\textbf{$@0.5$}&\textbf{$@1$}&\textbf{$@2$}&\textbf{$@4$}&Avg.[0.5,1,2,4]\\
\toprule[1pt]
Faster R-CNN \cite{ren2015fasterrcnn}&100\%&3&57.17 & 68.82 &74.97 &82.43&70.85\\
Faster R-CNN+SATr &100\%&3&62.52 (\textbf{5.35$\uparrow$}) & 73.91 (\textbf{5.09$\uparrow$}) &79.40 (\textbf{4.43$\uparrow$}) &86.03  (\textbf{3.60$\uparrow$})&75.47(\textbf{4.62$\uparrow$}) \\
\midrule
Faster R-CNN+cBM\cite{li2021conditional} &100\%&3&65.37  & 76.31  &81.03  &87.98  &77.67 \\
Faster R-CNN+cBM+SATr &100\%&3&67.41 (2.04$\uparrow$) & 78.02 (1.71$\uparrow$) &82.43 (1.40$\uparrow$) &88.90  (0.92$\uparrow$)&79.19(1.52$\uparrow$) \\
\midrule
3DCE  \cite{yan20183DCE} &100\%&9 &59.32 &70.68&79.09&84.34&73.36\\
3DCE+SATr &100\%&9&64.38 (\textcolor[rgb]{1,0,0}{5.06$\uparrow$})&75.55 (\textcolor[rgb]{1,0,0}{4.87$\uparrow$}) &82.74  (\textcolor[rgb]{1,0,0}{3.65$\uparrow$})& 87.78 (\textcolor[rgb]{1,0,0}{3.44$\uparrow$})&77.61(\textcolor[rgb]{1,0,0}{4.25$\uparrow$})\\
\midrule
3DCE+cBM\cite{li2021conditional} &100\%&9&66.98&77.25 &83.64 & 88.41&79.07\\
3DCE+cBM+SATr &100\%&9&68.12 (1.14$\uparrow$)&78.33 (1.08$\uparrow$) &84.57  (0.93$\uparrow$)& 89.21 (0.80$\uparrow$)&80.06(0.99$\uparrow$)\\
\midrule

MVP-Net  \cite{li2019mvp}&100\%&9&70.07&78.77&84.91&87.33&80.27\\
MVP-Net+SATr&100\%&9&72.34 (2.27$\uparrow$)&80.27 (1.50$\uparrow$) &86.11 (1.20$\uparrow$)& 88.21 (0.88$\uparrow$)&81.73(1.46$\uparrow$)\\
\midrule
MVP-Net+cBM\cite{li2021conditional}&100\%&9&73.05 &81.41  &87.22 & 89.37 &82.76\\
MVP-Net+cBM+SATr&100\%&9&74.11 (1.06$\uparrow$)&82.31 (0.90$\uparrow$) &88.14 (0.92$\uparrow$)& 90.13 (0.76$\uparrow$)&83.67(0.91$\uparrow$)\\

\midrule
AlignShift \cite{yang2020alignshift}&100\%&7&77.20&84.38&89.03&92.31&85.73\\
AlignShift+SATr&100\%&7&78.98 (1.78$\uparrow$)&85.82 (1.44$\uparrow$) &90.21 (1.18$\uparrow$)& 93.27 (0.96$\uparrow$)&87.07(1.34$\uparrow$)\\

\midrule
AlignShift+cBM\cite{li2021conditional} &100\%&7&79.17 &85.71 &89.80&92.65&86.83\\
AlignShift+cBM+SATr&100\%&7&\textcolor[rgb]{1,0,0}{79.98} (0.81$\uparrow$)&\textcolor[rgb]{1,0,0}{86.36} (0.65$\uparrow$) &\textcolor[rgb]{1,0,0}{90.22} (0.42$\uparrow$)& \textcolor[rgb]{1,0,0}{92.99} (0.34$\uparrow$)&\textcolor[rgb]{1,0,0}{87.39}(0.56$\uparrow$)\\
\midrule
A3D w/o Fusion\cite{yang2021asymmetric}&100\% &7&72.47&81.35&86.68&90.41&82.73\\
A3D w/o Fusion +SATr&100\% &7&74.68(2.21$\uparrow$)&83.17(1.82$\uparrow$)&88.24(1.56$\uparrow$)&91.58(1.17$\uparrow$)&84.42(1.69$\uparrow$)\\
\midrule
A3D\cite{yang2021asymmetric} &100\%&7&79.24&85.04&89.15&92.71&86.54\\
A3D+SATr&100\%&7&\textbf{81.03} (1.79$\uparrow$)&\textbf{86.64} (1.60$\uparrow$) &\textbf{90.70} (1.55$\uparrow$)& \textbf{93.30}( 0.59$\uparrow$)&\textbf{87.92}(1.38$\uparrow$)\\

\bottomrule[1.5pt]
\end{tabular}

\end{lrbox}
\scalebox{0.6}{\usebox{\tablebox}}
\end{table}
\subsection{Lesion detection performance}
Five state-of-the-art ULD approaches \ucite{yan20183DCE,li2019mvp,yang2020alignshift,yang2021asymmetric,li2021conditional} and one natural image  \ucite{ren2015fasterrcnn} detection method are compared to evaluate SATr's effectiveness.

\textbf{Full training dataset results.} As shown in Table \ref{results},  our method brings promising detection performance improvements for all baselines with full training dataset. The improvements of Faster R-CNN \cite{ren2015fasterrcnn}, 3DCE, 3DCE w/ cBM  and MVP-Net are more pronounced than those of AlignShift \cite{yang2020alignshift} and A3D \cite{yang2021asymmetric}. This is because AlignShift and A3D introduce channel-fusion mechanism among different slices in backbone, thus the v value enhancement design in SATr brings less advances. Anyway, SATr still endows A3D and AlignShift with strong global dependency catching power to provide notable improvements. Meantime, the A3D w/ SATr reached the SOTA result. 

\textbf{Partial training dataset results.} As shown in Table \ref{results}, under the $25\%$ and $50\%$ training data settings, our method brings more improvements than that in the full training data scene. Although the bigger performance improving space plays a huge role, we can also attribute this to our value removal design. When the training data is limited, SATr forces the network to learn more dependency to avoid overfitting. Besides, our experiments also  show some results contradicted with the full-training dataset experiment, e.g., AlignShift outperformed A3D.

\textbf{Classification activation maps results.}
We showcase CAMs (based on \cite{jacobgilpytorchcam} \cite{muhammad2020eigen}) of two SOTA ULD methods to evaluate the global context modeling power of SATr. As shown in \ref{fig:fig1_CAM}, SATr helps model useful dependency  within single slice ($a_1$v.s.$A_1$ and $a_2$v.s.$A_2$) and among different slices ($b_1$v.s.$B_1$ and $b_2$v.s.$B_2$). More cases are included in Suppl. Material.
\begin{table}[t]
\centering
\caption{Sensitivity (\%) at various FPPI on the testing dataset of DeepLesion \cite{yan18deeplesion} under 25\%  and 50\% training data settings.} \label{results}
\begin{lrbox}{\tablebox}
\begin{tabular}{p{40mm}p{13mm}<{\centering}p{12mm}<{\centering}p{23mm}p{23mm}p{23mm}p{23mm}p{23mm}}
  \toprule[1.5pt]
  \textbf{Methods}&\textbf{data}&\textbf{slices}&\textbf{$@0.5$}&\textbf{$@1$}&\textbf{$@2$}&\textbf{$@4$}&Avg.[0.5,1,2,4]\\
\toprule[1pt]

AlignShift\cite{yang2020alignshift}&25\% &7&52.17&62.17&69.50&75.24&64.77\\
AlignShift+SATr&25\% &7&56.68(4.51$\uparrow$)&65.04(2.87$\uparrow$)&72.03(2.53$\uparrow$)&77.83(\textbf{2.59$\uparrow$})&67.90(3.13$\uparrow$)\\

\midrule
 AlignShift+cBM\cite{li2021conditional} &25\% &7&56.84&64.96&71.86& 81.08&68.69\\
 AlignShift+cBM+SATr&25\% &7&\textbf{62.31} (\textbf{5.47$\uparrow$})&\textbf{70.13} (\textbf{5.17$\uparrow$}) &\textbf{76.79} (\textbf{4.93$\uparrow$})& \textbf{81.17} (0.09$\uparrow$)&\textbf{72.60}(\textbf{3.91$\uparrow$})\\

\midrule
A3D\cite{yang2021asymmetric}&25\% &7&55.67&65.39 &73.35& 79.31&68.43\\
A3D+SATr&25\% &7&59.99 (4.32$\uparrow$)&68.05 (2.66$\uparrow$) &74.67 (1.32$\uparrow$)& 79.09 (0.22$\downarrow$)&70.45(2.02$\uparrow$)\\

\bottomrule[1.5pt]
AlignShift\cite{yang2020alignshift}&50\% &7&68.69&76.73&82.25&86.54&78.55\\
AlignShift+SATr&50\% &7&71.10(2.41$\uparrow$)&78.32 (1.59$\uparrow$) &83.51 (1.26$\uparrow$)& 87.99 (\textbf{1.45$\uparrow$})&80.23(1.68$\uparrow$)\\

\midrule
 AlignShift+cBM\cite{li2021conditional} &50\%&7&70.30&78.16&83.35&87.99&79.95 \\
 AlignShift+cBM+SATr&50\% &7&73.53 (\textbf{3.23$\uparrow$})&79.91 (1.75$\uparrow$) &84.89 (\textbf{1.54$\uparrow$})& 88.50 (0.51$\uparrow$)&81.71(\textbf{1.76$\uparrow$})\\

\midrule
A3D\cite{yang2021asymmetric} &50\% &7&72.52&80.27&86.14&90.15&82.27\\
A3D+SATr&50\% &7&\textbf{75.24} (2.72$\uparrow$)&\textbf{82.19} (\textbf{1.92$\uparrow$}) &\textbf{86.99} (0.85$\uparrow$)& \textbf{90.96} (0.81$\uparrow$)&\textbf{83.85}(1.58$\uparrow$)\\

\bottomrule[1.5pt]
\end{tabular}

\end{lrbox}
\scalebox{0.6}{\usebox{\tablebox}}
\end{table}

\subsection{Ablation study}
An ablation study is provided to evaluate the importance of the two key designs: (i) Enhancement of value vector with key-slice feature and (ii) Removal of key-slice feature from the query and key vectors. Adding naive transformer block for baseline, the performance is increased by 0.44\%, enhancing the value vector with the key-slice feature, we obtain a 0.94\% improvement over the naive transformer block. Further removing the key-slice feature from the query and key vectors accounts for another 0.83\% improvement and give the best performance.
Ablation study for different SATr blocks, which is included in Suppl. Material, shows that  all SATr blocks is important.

\begin{table}[t]
\centering
\caption{Ablation study of our method at various FPs per image (FPPI).} 
\label{ablation_study}
\begin{lrbox}{\tablebox}
\begin{tabular}{p{24mm}<{\centering}p{26mm}<{\centering}p{24mm}<{\centering}p{24mm}<{\centering}p{22mm}p{22mm}}
\toprule
A3D~\cite{yang2021asymmetric} w/o fusion&Naive transformer block&Enhancement of value vector &Removal of key-slice feature &$FPPI$=0.5&$FPPI$=1\\ \hline
$\checkmark$&&&&72.47&81.35\\
$\checkmark$&$\checkmark$&&&72.91(0.44$\uparrow$)&81.68(0.33$\uparrow$)\\
$\checkmark$&$\checkmark$&$\checkmark$&&73.85(\textbf{0.94}$\uparrow$)&82.48(\textbf{0.80}$\uparrow$)\\
$\checkmark$&$\checkmark$&$\checkmark$&$\checkmark$&\textbf{74.68}(0.83
$\uparrow$)&\textbf{83.17}(0.69$\uparrow$)\\

\bottomrule
\end{tabular}

\end{lrbox}
\scalebox{0.7}{\usebox{\tablebox}}
\end{table}
\section{Conclusion}
Multi-slice-input based ULD methods using CNN backbones have inherent limitations in capturing the global contextual information within individual slice and among multiple adjacent slices. To address this issue, we propose a slice attention transformer (SATr) block that can be integrated with conventional CNN backbones to obtain better global representation.
Extensive experiments using several SOTA ULD methods as baselines show that the proposed can  be easily integrated with many existing methods to boost their performance without extra hyperparameters or special network designs.

\newpage
\bibliographystyle{splncs04}
\bibliography{egbib}

\end{document}